\documentclass[10pt,letterpaper]{article}
\usepackage[top=0.85in,left=2.75in,footskip=0.75in]{geometry}

\usepackage{changepage}

\usepackage[utf8]{inputenc}

\usepackage{textcomp,marvosym}

\usepackage{fixltx2e}

\usepackage{amsmath,amssymb}

\usepackage{cite}

\usepackage{nameref,hyperref}

\usepackage[right]{lineno}

\usepackage{microtype}
\DisableLigatures[f]{encoding = *, family = * }

\usepackage{rotating}

\raggedright
\usepackage[aboveskip=1pt,labelfont=bf,labelsep=period,justification=raggedright,singlelinecheck=off]{caption}

\makeatletter
\renewcommand{\@biblabel}[1]{\quad#1.}
\makeatother

\date{}

\usepackage{lastpage,fancyhdr,graphicx}
\usepackage{epstopdf}
\fancyheadoffset[L]{2.25in}
\fancyfootoffset[L]{2.25in}

\begin{document}

\vspace*{0.35in}

\begin{flushleft}
{\Large
\textbf\newline{The Dynamics of Initiative in Communication Networks}
}
\newline
\\
Anders Mollgaard\textsuperscript{},
Joachim Mathiesen\textsuperscript{*}\\
\bigskip
University of Copenhagen, Niels Bohr Institute , 2100 Copenhagen, Denmark\\
\bigskip
* mathies@nbi.dk

\end{flushleft}

\section*{Abstract}
Human social interaction is often intermittent. Two acquainted persons can have extended periods without social interaction punctuated by periods of repeated interaction. In this case, the repeated interaction can be characterized by a seed initiative by either of the persons and a number of follow-up interactions. The tendency to initiate social interaction plays an important role in the formation of social networks and is in general not symmetric between persons. In this paper, we study the dynamics of initiative by analysing and modeling a detailed call and text message network sampled from a group of 700 individuals. We show that in an average relationship between two individuals, one part is almost twice as likely to initiate communication compared to the other part. The asymmetry has social consequences and ultimately might lead to the discontinuation of a relationship. We explain the observed asymmetry by a positive feedback mechanism where individuals already taking initiative are more likely to take initiative in the future. In general, people with many initiatives receive attention from a broader spectrum of friends than people with few initiatives. Lastly, we compare the likelihood of taking initiative with the basic personality traits of the five factor model. 

\section*{Introduction}
The digital recording of our social life including telecommunication and activity on the internet allows us to answer previously unanswered questions about human behavior. It is now possible to study the complex dynamics of social networks \cite{holme2012temporal, karsai2011small} as well as their formation and stability \cite{albert2002statistical, newman2003structure, kossinets2006empirical, oh2007membership, barabasi1999emergence}. Here, we are particularly interested in the dynamics of initiative in social relations and how it might influence the formation of social networks.
A high degree of asymmetry in relations is observed in phone call networks \cite{kovanen2010reciprocity} simply by counting the number of calls going in each direction for all links over a period of time. In particular, it has been found that for 25\% of the links, 80\% or more of the calls in a relationship come from one part alone. Also, it has been shown that this asymmetry cannot be explained by personal differences in call frequency. In general, it is possible to take initiative through various modes of interaction, e.g.~by face-to-face meetings, which might result in a later call by phone or a text message. The fact that communication happens in different ways might lead to spurious estimates of the asymmetry in relations, when computed from a single mode of communication.


Here we consider communication in two channels, phone calls and text messages, in a detailed data set sampled from 700 individuals. The data is collected from smartphones distributed to students at the Technical University of Denmark and includes information about communication both inside and outside the group of students. The data is recorded over a period of 83 weeks and consists of 32412 links based on 972592 calls and 3949710 text messages. Smartphones have previously been used to infer 95\% of self-reported friendships based on communication and proximity \cite{eagle2009inferring}. In addition to this, mobile phones have been used in several studies on human behavior \cite{lane2010survey, stopczynski2014measuring} including communication dynamics \cite{hidalgo2008dynamics}, mobility \cite{song2010limits}, and personality \cite{de2013predicting,mollgaard2016homophily}.

In the following, we introduce a parameter describing asymmetry in social relations. From a statistical model that takes into account the sampling error on individual links, we will compute the most probable distribution of the asymmetry parameter. We further consider the dynamics of initiative by analysing the implications and causation of asymmetry in relations. 

\section*{Results}
\begin{figure*}[h!]
	\centering
	\includegraphics[width=0.6\textwidth]{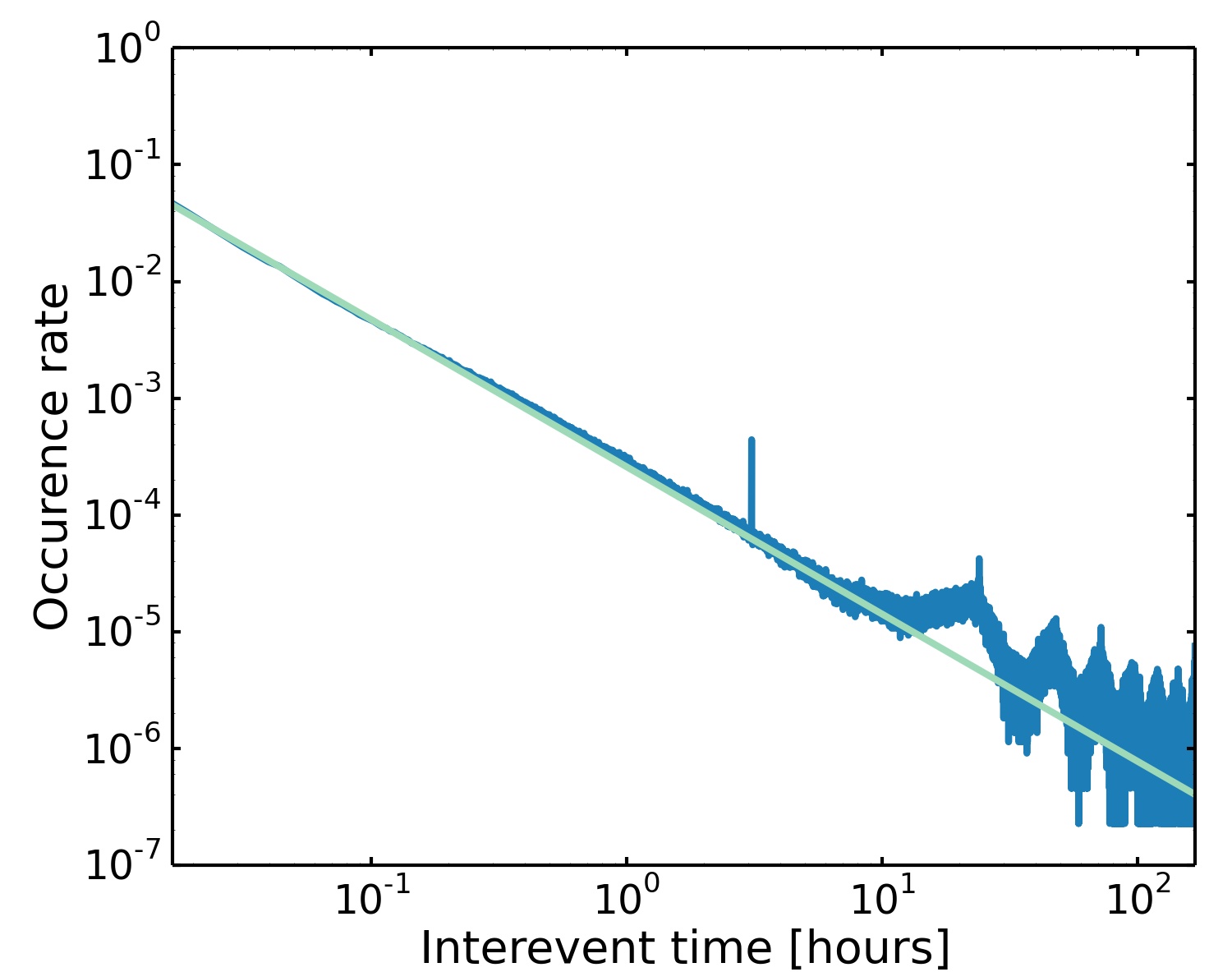}
	\caption{\textbf{Interevent distribution.} Log-log plot of the distribution of waiting times between communication events including both calls and text messages. Also shown is a power law with an exponent of $\alpha = -1.26$, which fits the distribution all the way from one minute to one week. The power law nature of the distribution suggests that there is no natural time scale to separate initiatives from follow-up communication. The local peak at 3 hours is possibly the electronic response of an app and the oscillations at large inter event times correspond to circadian cycles.}
	\label{fig:inter event}
\end{figure*}

\textbf{Definition of initiative.} Not all phone calls and text messages represent an initiative, since many of them are responses to previous communication. For example, a call may not be picked up, but will anyway result in a later answer by the called part. Likewise, most text messages will be part of an on-going communication. In order to distinguish initiative from follow-up communication, we need to define a time scale that separates the two. In Fig.~\ref{fig:inter event}, we show the distribution of times between communication events over all relations in the data set. The time axis spans from one minute to one week and the distribution follows a power law with an exponent of $\alpha = -1.26$. The oscillations at large time scales are due to the circadian rhythm and there is a local peak at precisely 3 hours, which is possibly due to the electronic response of an app installed on a number of the smartphones in the study. The observed power law is in line with previous results for inter-event times of human communication \cite{barabasi2005origin, malmgren2008poissonian, vazquez2007impact, stouffer2006log, nakamura2007universal} and other human activity \cite{johansen2000download, stouffer2006log}. Unfortunately this also means that there is no natural time scale to separate initiative from follow-up communication. Nonetheless, in this analysis, we have chosen to make a cut between follow-up interaction and initiative at 24 hours. Most on-going telephone communication has inter-event times below this scale (88 \% of all inter-event times), such that the expected rate of false positives for initiatives will be rather low. We therefore define an initiative to be a communication event that is separated from a prior communication on the same link by more than 24 hours. Communication separated by a shorter time is considered to be follow-up.

\textbf{Model.} In the following, we compute the maximum-likelihood estimate of the asymmetry distribution for relationships based on a statistical model. The relationship between two persons, $A$ and $B$, consists of a series of initiatives. For simplicity, we shall assume that the probability for an initiative to be taken by person $A$, $\gamma_A$, is constant over time. The corresponding probability for $B$ is then $\gamma_B = 1 - \gamma_A$. We introduce $\mu \equiv \textrm{min} \left( \gamma_A , \gamma_B \right) \le 0.5$, as a measure of the asymmetry in a relationship. $\mu = 0.5$ represents a fully symmetric relation with respect to initiatives, while $\mu = 0$ represents a fully asymmetric one. In the following, we will estimate the distribution of $\mu$ across relationships in the full social network, which we shall denote $F \left( \mu \right)$. While we cannot compute the true distribution, we shall compute the maximum-likelihood distribution, $F_{\textrm{ML}} \left( \mu \right)$, based on the above assumptions. 

Our data consists of a call and text message network of participants in the study and external people that the participants have communicated with. The  $M$ observed links in the network form a set, $D = \left\lbrace X_i | i=1,...,M \right\rbrace$, where $i$ enumerates each link. The links $X_i \equiv \left(n_{A,i},n_{B,i}\right)$ are characterized by the number of initiatives from person $A$ ($n_{A,i}$) and $B$ ($n_{B,i}$), respectively, and we define $N_{AB,i}=n_{A,i}+n_{B,i}$.

Invoking Bayes' theorem we get

\begin{align*}
P \left( F | D \right) &= \frac{P \left( D | F \right) P \left( F \right)}{P \left( D \right)} , \\
&\propto P \left( D | F \right), \\
&= \prod_i P \left( X_i | F \right).
\end{align*}
For simplicity, we have here assumed a constant prior over $F$ along with statistical independence of the relationships. The normalization $P \left( D \right)$ is ignored, since it is a constant with respect to variation in $F$. The least initiative taking person could be person $A$ or person $B$, we must therefore integrate over both possibilities. Let us denote the least initiative taking person by $Y$. The likelihood factor associated with a link $\left(n_{A,i},n_{B,i}\right)$ given $\mu$ and $Y$ may then be calculated using the binomial distribution

\begin{align*}
P \left( X_i | F \right) &= \sum_{ Y \in \left\lbrace A,B \right\rbrace } \int P \left( X_i | Y, \mu \right)  P \left( Y \right) F \left( \mu \right) \, d\mu, \\
&= \sum_{ n_i \in \left\lbrace n_{A,i}, n_{B,i} \right\rbrace } \int_{0}^{\frac{1}{2}} \binom{N_i}{n_i} \mu^n_i \left( 1 - \mu \right)^{N_i-n_i} \cdot \frac{1}{2} \cdot F \left( \mu \right)  \, d\mu.
\end{align*}
Here we sum over $Y$ (or equivalently $n$) giving each of the two possibilities a prior of a ${1}/{2}$ and likewise integrate over $\mu$ giving each value a prior of $F\left( \mu \right)$. We shall now find the distribution $F_{\textrm{ML}} \left( \mu \right)$ that maximizes the likelihood function $P \left( D | F \right)$, or equivalently, the log-likelihood $\log P \left( D | F \right) $. 

We do this numerically using sequential least squares programming \cite{kraft1988software}, which handles both constraints and bounds. In our statistical model, we discretize the initiative parameter $\mu \rightarrow \mu_j$ and thereby we also discretize the distribution $F \left( \mu \right) \rightarrow F_{j}$. The constraint $\sum_j F_{j} = 1$ and the bounds $0 \le F_{j} \le 1$ are introduced. We initialize $F_{j}$ according to the distribution of $n_{A,i} / N_i$ restricted to $N_i > 20$ and let the sequential least squares estimation run until convergence. The analytic expression for the gradient is calculated and used in the optimization, and for $N > 100$, we approximate the binomial distribution by the normal distribution in order to handle the product between the very large $\binom{N_i}{n_i}$ and the very small $\mu^n_i \left( 1 - \mu \right)^{N_i-n_i}$. 

In order to test the robustness and convergence of our estimates, we have created 100 synthetic data sets in the following way. We keep the number of links identical to the original data set. Similarly, we keep the total number of initiatives, $N_{AB,i}$, on each link, $X_{i}$, the same, but we disregard the direction of the initiatives. For each link, we then draw a $\mu$ from the distribution $F_{\textrm{ML}}(\mu)$, which is then used in a stochastic simulation of the direction of the $N_{AB,i}$ initiatives. From the synthetics data set, we then make a maximum likelihood estimate estimate, $\tilde{F}_{\textrm{ML}}(\mu)$ of the underlying distribution $F_{\textrm{ML}}(\mu)$.

\begin{figure*}[h!]
	\centering
	\includegraphics[width=0.55\textwidth]{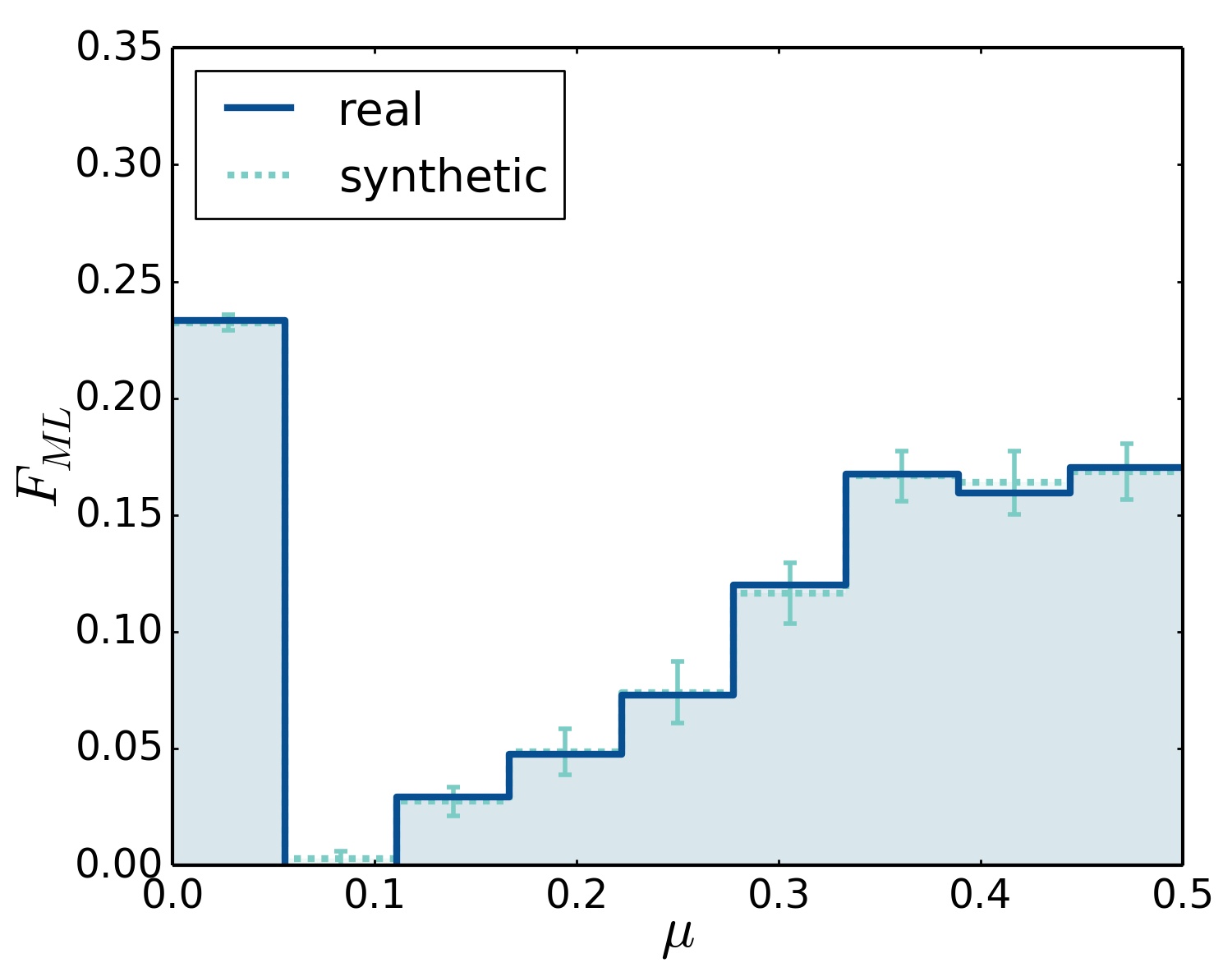}
	\caption{\textbf{Asymmetry in relations.} We show the max likelihood distribution of the initiative parameter across relationships. Here $\mu = 0.5$ corresponding to a fully symmetric relationship, while $\mu = 0$ is a fully asymmetric one. The full line corresponds to the max likelihood distribution as it is derived from the data. We also test the method for bias and uncertainty by applying it to synthetic data sets. The dashed line correspond to the average estimate among the synthetic data sets, and the error bars correspond to the spread in these estimates. We suspect that many of the highly asymmetric relations at $\mu = 0$ are of non-social character. }
	\label{fig:relations}
\end{figure*}

In Fig.~\ref{fig:relations}, we show the max likelihood distribution of $\mu$ obtained for the real data set (full), along with the average distribution obtained for the synthetic data sets (dashed). Also shown is the spread in the results of the synthetic data sets. Note that our estimate of the underlying distribution is unbiased (the synthetic and real data overlap). We observe that the distribution is bimodal with 23 percent of the links being fully asymmetric, i.e.\ $\mu$ being very small. We suspect that many of these relations are of a less social character, e.g.~automatic text-message updates from service providers. For this reason, we restrict our analysis to reciprocal relations only in the following sections, i.e.~relations with at least one initiative in both directions. For the full distribution, the average $\mu$ is $0.282 \pm 0.001$, but if we exclude the potentially non-social links at $\mu = 0$, then we get an average of $0.359 \pm 0.001$. The errors correspond to those associated with the error of the mean in the simulated data sets. We conclude that in an average relation, one of the persons will be almost twice as likely as the other to initiate communication. 

\textbf{Dynamics of relationships.} In the above we considered relationships from a static point of view and assumed that the probability parameter was constant in time. Here we acknowledge that the history of a relationship may have an impact on future communication. In particular, we study the effects of ``initiative length''. To explain this term, we consider the following initiative history between person $A$ and person $B$: $[\ldots,B,A,A,A]$. The current initiative length is 3, because person A has taken 3 consecutive initiatives, since the last initiative taken by person B. In other words, the initiative length denotes the number of consecutive initiatives taken, since the last change of ``direction'' in initiative.

As the initiative length grows, i.e.~as either person $A$ or person $B$ keeps approaching the other without a turn of initiative, we expect that the relationship becomes more fragile and possibly comes to an end	. Discontinued relationships are difficult to identify in the data, since they might be resumed subsequent to the data recording. Instead, we look for relationships that have been inactive for a very long time. In particular, we consider the relationship between person $A$ and person $B$ to be discontinued, if the following is true: person $A$ has taken a total number of initiatives with other individuals, not including $B$, which is 10 times larger, than the average number of initiatives previously separating interaction between $A$ and $B$. 

\begin{figure*}[h!]
	\centering
	\includegraphics[width=0.47\textwidth]{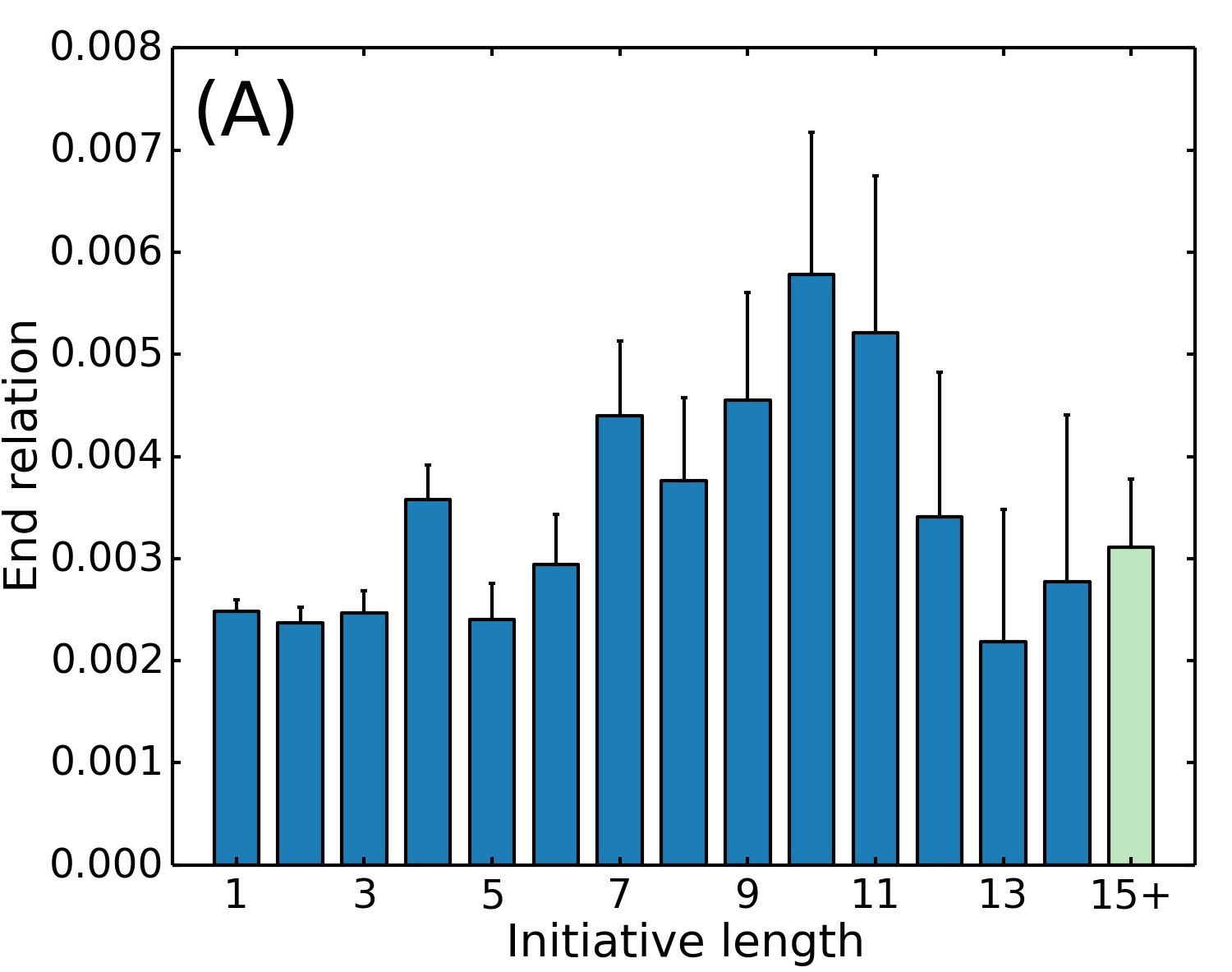}
	\includegraphics[width=0.47\textwidth]{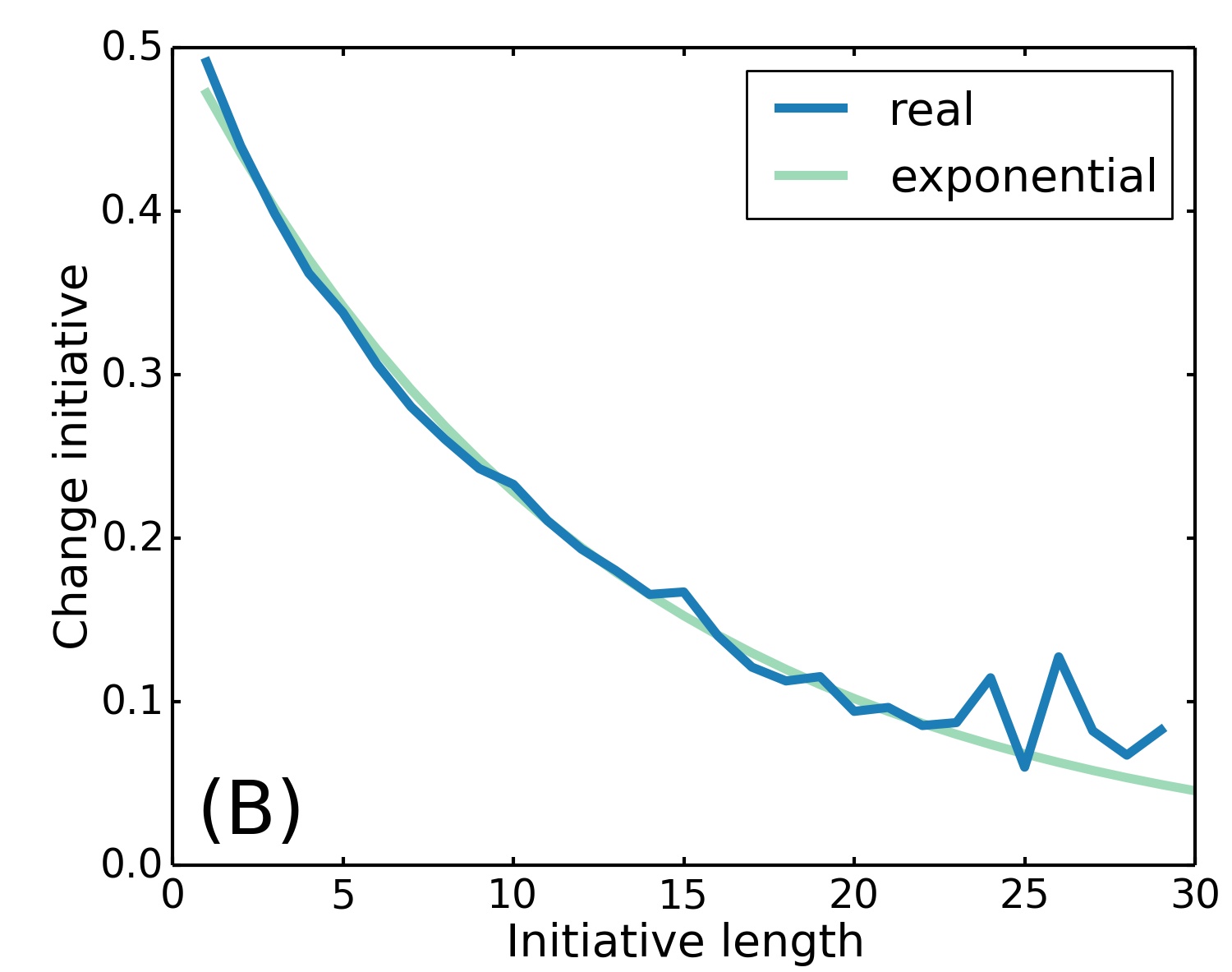}
	\caption{\textbf{Initiative dynamics.} In (A) we show the probability of a relationship ending after a given initiative length. Here ``initiative length'' denotes the number of consecutive one sided initiatives. Note that relationships are more likely to end after several one-sided initiatives, although the probability drops back again for very large (13+) initiative lengths. In (B) we show the probability of the initiative changing direction as a function of the initiative length. We see that the probability drops exponentially as the number of one sided initiatives in a row is increased. This suggests that initiatives promotes the role of an initiator through a positive feedback. }
	\label{fig:relations2}
\end{figure*}

In Fig.~\ref{fig:relations2}A we show the probability of a relationship ending as a function of the initiative length. We find that the probability generally increases as the initiative length grows towards 10. This is in line with previous research \cite{hidalgo2008dynamics}, which has shown that reciprocity is important to link persistence. For initiative lengths greater than 12, the probability drops back again; possibly because this involves only the most dedicated relations, e.g.\ close relatives. In general, we find among all relations that the average initiative length prior to the discontinuation of a relation is $3.2$ (which in a jargon could be named the ``ghosting factor''). Note that, for better statistics, we have restricted the analysis to relations with at least 15 initiatives. 

In Fig.~\ref{fig:relations2}B, we show that the probability for the initiative in a relationship to change from one part to the other as function of the initiative length. The plot is based on the statistics of all relationships. When person $A$ has taken a single initiative in a relationship between $A$ and $B$ there is a 50\% chance that the next initiative is taken by $B$. However, as the number of consecutive initiatives taken by $A$ increases, the probability for $B$ to take the next initiative drops exponentially. A least square exponential fit yields the relation $y = 0.51\cdot  0.92^{x}$, which says that the chance of the initiative changing direction drops by 8\% each time you make an initiative. In other words, the direction of an initiative seems to have a self-amplifying effect, which promotes the role of an initiator. This mechanism provides a possible explanation for the persistent asymmetry observed in the data.

\begin{figure*}[h!]
	\centering
	\includegraphics[width=0.47\textwidth]{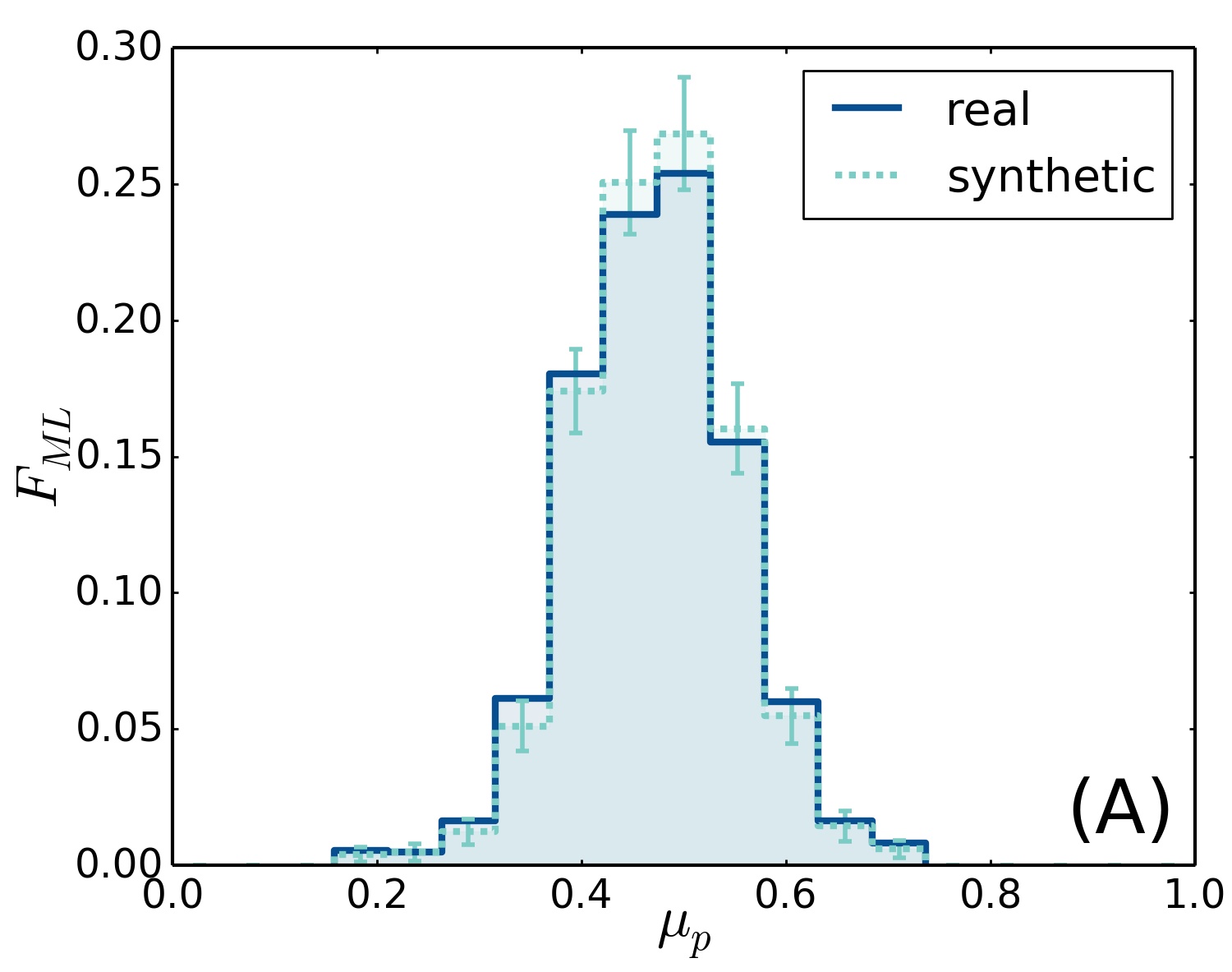}
	\includegraphics[width=0.47\textwidth]{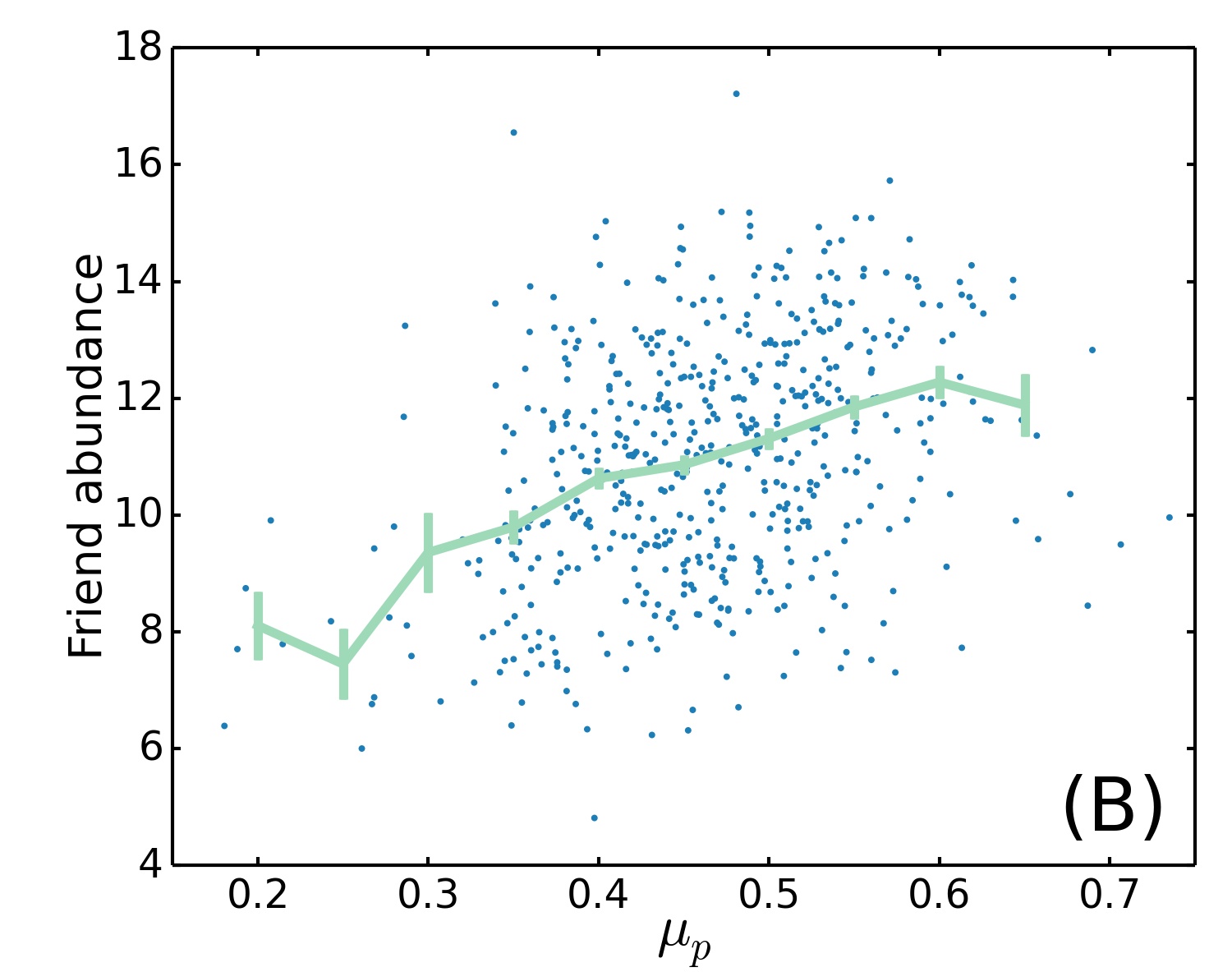}
	\caption{\textbf{Initiative statistics of individuals.} In (A) we show the maximum-likelihood distribution of the personal initiative parameter $\mu_p$, where $\mu_p$ is the probability that an initiative involving person A is attributed to him rather than to his friends. We note that $\mu_p$ varies a lot over the population; some people making only 25\% of the initiatives themselves against 70\% at the other end of the spectrum. This has social consequences. In (B) we have estimated $\mu_p$ and plotted it against the ``friend abundance'' for the full population. Friend abundance is estimated as the average number of unique friends among all consecutive combinations of 20 incoming initiatives. We find that people with large $\mu_p$ are rewarded by increased attention from their network. }
	\label{fig:persons}
\end{figure*}

\textbf{Dynamics of individuals.} 
 Here we turn to the initiative statistics of a person rather than a relationship. We suspect that people, likely to take many initiatives, are more likely to make friends, and, as the above results suggest, they are also more likely to keep them. Again we start by deriving the population statistics of the initiative parameter $\mu_p$, only this time it characterizes a person instead of a link and it runs from 0 to 1. The parameter measures the probability that an initiative involving person $i$ is taken by $i$ rather than one of her/his friends. Each person is characterized by the number of in- and outgoing initiatives across all her/his relations. The analysis follows the same steps as above, but without the sum over $Y$ and with the integral over $\mu_p$ running from 0 to 1. The obtained maximum-likelihood distribution for the $\mu_p$ parameter is shown in Fig.~\ref{fig:persons}A. The distribution is a little biased towards low initiative with a mean of $0.469 \pm 0.003$, possibly because students are more likely to be called by their parents than the other way around. The spread of the distribution is $0.082 \pm 0.002$, but this estimate is probably slightly biased towards 0, due to the fully asymmetric links being neglected. This is obvious from the spread of the reference distributions (again computed from synthetic data), which is 4\% lower than the spread of the distribution from which it was simulated. In general, the few percent with the very lowest $\mu$ only take the initiative 25\% of the time, whereas in the other end, we find individuals which take initiative in 70\% of the cases. 
 
Next we look at the social consequences of $\mu_p$; specifically we wish to quantify the impact of a person's initiative on his abundance of friends. Since communication has been recorded for varying lengths of periods, we cannot simply count the number of contacts. Instead we count the average number of unique friends among all groups of 20 consecutive incoming initiatives. In principle, all 20 initiatives could be attributed to a single person in which case we count just 1 unique friend among the 20 initiatives. The other end of the scale corresponds to the 20 initiatives being attributed to 20 different people, i.e. 20 unique friends. If one counts many unique individuals among person A's incoming initiatives, it is an indication that many different people take an interest in her/him. The $\mu_p$ parameter of a person is estimated from the number of outgoing initiatives relative to the total number of initiatives. A scatter plot of the two quantities is shown in Fig.~\ref{fig:persons}B and we find a correlation between the two numbers of $0.35 \pm 0.04$. The mean value of a binning along the horizontal axis is also shown in the plot. The mean shows a clear trend: more personal initiative implies more interest from friends. In particular, the number of unique friends among 20 incoming initiatives jumps from 8 to 12 as we move from low to high personal initiative. Note that, for better statistics, this analysis has been restricted to individuals involved in at least 200 initiatives.

Finally we compare the likelihood of taking initiative with basic personality traits. The participants in the study have filled questionnaires providing information about the Big Five personality traits \cite{Digman1990a}. We have then performed a linear correlation between the initiative parameter and the personality scores for each individual. We find the following correlation coefficients for the traits with errors given as one standard deviation, agreeableness (0.02 $\pm$ 0.12 ), conscientiousness (0.13 $\pm$ 0.10), extraversion (0.26$\pm$0.06), neuroticism (0.11$\pm$0.07) and openness (0.07$\pm$0.08). The trait with the highest correlation coefficient is perhaps not surprisingly extraversion. In general, people with high extraversion scores are described to be outgoing and have a high degree of sociability.

\section*{Discussion}

The individual initiative for social interaction is an important factor in the formation and stability of social networks. 
Here we have applied a statistical model to derive the distribution of an initiative parameter in the social network of 700 individuals. In an average relationship, we find that one part will take initiative only 36\% of the time whereas the other part will take initiative the remaining 64\%, i.e.\ one part is in general twice as likely to take initiative as the other. These numbers represent a considerable asymmetry among the bulk of all relationships. We find that relationships are more likely to end after large initiative lengths, although it drops back again around $13$ consecutive initiatives by one part. Despite the possible implications for the stability of a relationship, we observe no change in behavior to reduce the asymmetry. On the contrary, we find that the initiative bias is self-amplifying; i.e.\ the probability of an initiative changing side drops exponentially as a function of initiative length. The self-amplification provides a possible mechanism behind the observed asymmetry by promoting the role of an initiator over the history of a relationship.

We also derive the distribution of the initiative probability for individuals against all of their friends. Our population is found to be a little biased towards low initiative, with a mean value of 0.47 and a spread of 0.08. We show that persons with great personal initiative are rewarded by a greater abundance of incoming initiatives.
This suggests that the gain of taking initiative, e.g.\ making more friends and keeping them, is greater than the cost of being initiator. We also find that people taking many initiatives score higher on extraversion in personality tests.
People interact socially in many ways and since our data is restricted to telephone communication, we neglect initiatives made face-to-face and through online platforms. This might bias our results somewhat toward asymmetric relations. 

It would be interesting to explore further the effects of initiative in temporal networks. Our results indicate that nodes have different probabilities of activating their links. This has an effect on information spreading, since information will be quickly shared by some nodes, while others partly work as dead ends. Finally, initiatives are important in making new contacts and in keeping them, thereby being an important factor in network formation. 



\section*{Acknowledgements }
The study received funding through the UCPH 2016 Excellence Programme for Interdisciplinary Research. 
\section*{Methods}
This study was reviewed and approved by the appropriate Danish authority, the Danish Data Protection Agency (Reference number: 2012-41-0664).
The Data Protection Agency guarantees that the project abides by Danish law and also considers potential ethical implications. All subjects in the study gave written informed consent. 
\section*{Supporting Information}
\subsection*{S1 Text}
\label{S1_text}
{\bf Description of data.}  Short description of data used in the manuscript. 
\subsection*{S2 Text}
\label{S1_data}
{\bf Data file.} Data on initiatives of individuals included in the study.


\end{document}